1     **Cosmology with New Astrophysical Constants**

3     *Antonio Alfonso-Faus*

4     *E.U.I.T. Aeronáutica*

5     *Plaza cardenal Cisneros s/n*

6     *8040 Madrid, SPAIN*

8     **Abstract.** It is shown that Einstein field equations give two solutions for cosmology. The first one is the standard well known representative of the present status of cosmology. We identify it with the local point of view of a flat Universe with the values for the cosmological parameters $\Omega_k = 0$, $\Omega_\Lambda = 2/3$, $\Omega_m = 1/3$. The second one is a new one that we identify with a cosmic point of view, as given by free photons, neutrinos, tachyons and gravity quanta. We apply a wave to particle technique to find the matter propagation equation. Then we prove that all gravitational radii are constant, regardless of the possible time variations of the physical properties like the speed of light c, the gravitational constant G or the mass m of fundamental particles. We find two cosmological constants, $c^3/G$ and mc, with the condition that the field equations be derived from the action principle. With this result, and the integration of the Bianchi identity, we prove the existence of the two solutions for cosmology. We then validate Weinberg relation and prove that the speed of light c is proportional to the Hubble parameter H, a cosmological view. Thus, the cosmic observer (free photons and the like) are found to see accelerated photons with a numerical value that has been observed as an anomalous acceleration of the Pioneer 10/11 spacecrafts. An initial inflation is present which converts the Planck size into the present size of the Universe in about 140 initial tic-tac.

## 1. - INTRODUCTION

No significant time variations have been consistently observed in any of the fundamental quantities of physics. This may be due to an intrinsic, absolute, constancy of each physical property (mass, length etc.), or it may be due to the fact that we are measuring them using as a reference an arbitrary unit with the same dimensions and the same time variation. If this is the case we obtain a pure number, dimensionless, that is the ratio between the two quantities of the same physical property at the same instant of time. If each physical quantity is constant, we obtain their ratio as a constant number. If each quantity is time varying, with the same law for all of them, then we still get the same time-constant value: the ratio of the two quantities at any instant of time. In order to observe a time varying ratio it is required that one or the two quantities vary with time and with different laws. But it is not very reasonable to assume different cosmological time-laws, if any, for two quantities of the same physical property, with the same dimensions. Then, if they have to have the same time-law, the only way to observe time variations in their ratio is by comparing the two values at a different cosmological time one from the other. This plausible assumption is what happens with the observed red shift from distant galaxies, the Hubble observations in the 20´s.

The characteristic frequency of an emitted photon at the LAB can be taken as constant with time. This is due to the impossibility to observe time variations at the LAB, when we measure the same characteristic frequency of the same element at different times. Frequencies are the ratio of the speed of light c to the characteristic wavelength λ, which is proportional to the Compton wavelength of the element, ℏ/mc, or to the de Broglie wavelength, ℏ/mv. Now, no time variation is evident in the Planck's constant ℏ. Information on lifetimes of beta decays (which depend on a high power of ℏ, $ℏ^7$) from radioactivity implies that ℏ must be a true universal constant. As far as the linear momentum mc is concerned (or mv) it is also a universal constant, as we will prove. Then the sizes of particles given by their Compton wavelength (or their de Broglie wavelength) are universal constants. The same is true for the wavelength λ of emitted photons, which is proportional to the sizes.

If there were any change with time in the frequency of a photon at the LAB ($c/\lambda$) it should be due to time variations in the speed of light c. But no such time variations are observed locally. Why? The speed of light c is the ratio between the size of a fundamental particle and the time light takes to travel this distance. Since we prove that the sizes are universal constants, should the time taken for light to travel this distance also be a constant? This time duration is proportional to the tic-tac of the atomic clocks. And there is no way to compare the duration of two different tic-tacs if all the clocks run the same way. Then we may consistently take these clocks as uniformly measuring "time". Hence the speed of light is a constant at the LAB, and we may define it by a constant value, and we do it. Under this point of view the local frequencies of photons do not shift. They can be taken as constant references to be compared with photons coming from different distances in the Universe. When we do this we observe the Hubble red shift. But, what is this? If all photons from the same element are emitted at the same frequency everywhere in the Universe and at any time, the observed red shift must be due to something that happened during the time they travel from a distant galaxy to our Earth. Light may then get "tired" during the travelling, but this assumption has encountered strong and deep difficulties. The standard cosmological model takes c as a constant and explains the red shift as a lengthening of the wavelengths of the photons. Then this forces to postulate that the whole Universe is expanding to explain such a lengthening.

But there is an alternative way to explain the red shift. Believing in the strengthening of the photon wavelengths implies to expand the whole Universe. This is due to the belief in the uniformity of atomic clocks and the constancy of c. Alternatively we propose, and we will give strong theoretical reasons for this proposal, to believe in a non-expanding Universe, no wavelength strengthening, and assume that the atomic clocks do shift. In this case the speed of light is actually decreasing with cosmological time. All clocks must drift the same way: their tic-tac duration shifting proportionally to the age that the Universe has at the time the tic-tac takes place. This is something we will prove too. An expanding time takes the place of the expansion of space. We will have a non expanding Universe with an expanding "time".

We are facing here two alternatives. One is the standard one of expansion of the Universe with homogeneous clocks, and a constant speed of light. The other one is a new one with a spatially static Universe and an expanding interval of time built in all clocks, with a corresponding speed of light decreasing with time. As we will see, this is a change of role in the protagonists: the standard Hubble parameter, running inversely proportional to time with a constant speed of light, is here shown to be the same as the speed of light (except for a constant length factor), so that this speed is decreasing with cosmological time. The two alternatives represent two points of view that coexist: the local one and the cosmological one. Scientifically this has to be tested in a similar way that the Ptolemaic and Copernicus ideas were treated: The application of William of Ockham´s razor may something to say here too. And, as we will see, our approach nicely explains the Pioneer 10/11 anomalous acceleration, something that the standard model has not explained so far. At any rate our model derives from the field equations and its validity is the same as the Einstein general relativity theory.

We will first prove that all gravitational radii of any mass M, within its proper volume, are constant with time. This is a very important result. The gravitational radius of a mass M is given by $2GM/c^2$. This means that regardless of any possible time variation of G, M or c, and as long as we are within the proper volume of M, its gravitational radius is a universal constant. And this is a general relativistic result. This is then a new astrophysical constant. If the Universe is a black hole it has constant size, and of course this will be valid at a cosmological level.

Secondly we will prove that, in order to be able to derive Einstein's field equations from the Action Principle, two relations $c^3/G$ (a mass rate) and Mc (a linear momentum) must be universal constants, regardless of any possible time variation of G, M or c. This result is coherent with the previous one: the ratio of these two relations is just the gravitational radius

132  of M. We have then 3 new astrophysical constants in total, each one with
133  its own physical meaning. The mass rate in the Universe, M/t, is a constant.
134  So is its total momentum content Mc.

135  Thirdly we will integrate the Bianchi identity and find that only two
136  solutions are possible for the model Universes predicted by general
137  relativity. One giving an expanding Universe with zero pressure, and a
138  cosmological constant, and the other giving a static Universe, with a gas of
139  gravity quanta pressure and zero cosmological constant. The two solutions
140  coexist, one local the other at a cosmological level.

141  As a fourth step we will validate Weinberg's relation, something that
142  to our knowledge has not yet been done. This validation, and the use of the
143  constants we have defined, immediately gives a very surprising result: the
144  speed of light and the Hubble parameter are proportional. Then Weinberg's
145  relation is equivalent to say that the product $Gm^3$ is a universal constant.
146  This is also the result of the lunar laser ranging experiments.

147  Finally we will integrate the Einstein cosmological equations,
148  predicting the numerical values of some parameters that have already been
149  observed. Both, from a local point of view and from a "cosmic" point of
150  view.

151
152
153

## 2. - THE CONSTANCY OF ALL GRAVITATIONAL RADII

By using a wave to particle interaction technique presented elsewhere, Adams 1982 and 1983, Alfonso-Faus 1986, we derive a matter propagation equation where no prior assumption is made of any possible time variation for neither G nor c:

$$\left( (\frac{G}{c^2} N P^\mu)_{;\alpha} \right) P^\alpha = 0 \qquad (1)$$

Using $P^\mu = mU^\mu$ we get

$$U^\mu_{;\alpha} U_\alpha + \frac{d \ln\left(\frac{G}{c^2} mN\right)}{dt} U^\mu U^0 = 0 \qquad (2)$$

and contracting with $U_\mu$ we finally get

$$\frac{d \ln\left(\frac{G}{c^2} mN\right)}{dt} = 0 \quad i.e. \quad \frac{GmN}{c^2} = const \qquad (3)$$

This means that all the gravitational radii of masses mN inside their proper volume are time constant. The possible expansion or contraction of the Universe does not affect them. The standard approach of taking G, c, m and N as constants can now be relaxed to allow them to vary with time, within the constrains of the 2 new astrophysical constants that ensure the derivation of the Einstein's field equations.

## 3. - THE TWO UNIVERSAL CONSTANTS FROM THE ACTION PRINCIPLE

The Einstein's field equations can be derived from an action integral A, Weinberg 1972:

$$A = I_G + I_M = - c^3 / (16\pi G) \int R\,(g)^{1/2}\,d^4 x \;\; + \;\; I_M \tag{4}$$

where $I_M$ is the matter action and $I_G$ the gravitational term. Then the variation of the coefficients in the integrals must be zero to be able to get the field equations. One must have then

$$\frac{c^3}{16\pi G} = const. \tag{5}$$

From here it is seen that a time varying c must include a time varying G and vice versa. This constant has dimensions of mass over time. It is of the order of magnitude of the mass of the Universe divided by its age today. Also the ratio of Planck's mass to Planck's time has the same value. We are tempted to say that this constant rate of mass has universal physical meaning, a universal mass rate present at a cosmological scale as well as the Planck's scale.

The action for a free material point is

$$A = -mc \int ds \tag{6}$$

To preserve mechanics the momentum mc must be constant. Any theory must be special relativistic correct. Then the ratio of speeds v/c, that is present in special relativity, must be constant with time. No change with cosmological time has been observed related to this ratio. Hence mc = constant implies mv = constant, the constancy of linear momentum. Any time variation of c requires a time variation of m and vice versa.

## 4. - THE BIANCHI IDENTITY INTEGRATED

We have derived elsewhere, Belinchón and Alfonso-Faus 2001, the expression for the zero value of the right hand side of the Einstein's field equations

$$\nabla (G/c^4 \cdot T^{\mu\nu}) = 0 \quad (7)$$

which is

$$\frac{\rho'}{\rho} + 3(\omega+1)H + \frac{\Lambda' c^4}{8\pi G\rho} + \frac{G'}{G} - 4\frac{c'}{c} = 0 \quad (8)$$

where we have allowed for the possible time dependence of G, c and Λ. We consider the cosmological parameter Λ to be a real constant of integration so that it disappears from (8). Here ρ is the energy density. Using the equation of state p = w ρ, integration of (8) gives

$$\frac{G\rho}{c^4} R^{3(\omega+1)} = const \quad (9)$$

or equivalently

$$\frac{GM}{c^2} R^{3w} = const \qquad (10)$$

This result is a very important one derived from general relativity. Since all the gravitational radii are constants, we arrive at the conclusion that the possible Universes predicted by the Einstein's field equations have to obey the important result:

$$R^{3w} = const \qquad (11)$$

Hence we have only two alternatives: either the Universe is expanding, R(t), and w = 0 (no pressure term present in the equations) as the only solution, or the Universe has constant size R and w is not yet determined by this equation. In this case it turns out that the gravitational radius of the Universe being a constant equals its visible size with R = ct = constant = L ≈ $10^{28}$ cm. This is a coherent result with the next section where we prove that the speed of light is proportional to the Hubble parameter, c = HL. Since H varies as 1/t the speed of light is decreasing linearly with time, the product ct being the constant size of the Universe. The old problem of explaining the "coincidence" implied by the gravitational radius of the Universe being of the order of ct is completely solved here. We have to stress that this conclusion of c decreasing with time is a direct result of the Einstein's general relativity theory and, as such, it will drive as to a well-developed and self-consistent physical model of the Universe. For M the present mass of the Universe we have then

$$2GM/c^2 = R = ct = L = \text{constant} \approx 10^{28} \text{ cm}. \tag{12}$$

No length in the Universe is expanding nor contracting. The dynamo paradox, that could ideally be constructed to obtain work from a rod attached to a galaxy that is going away from us, just goes away.

There is a way to interpret Mach's principle: the rest energy of any mass m is equal to its gravitational potential energy due to the rest of the mass of the Universe:

$$GMm/R \approx mc^2 \quad \text{or} \quad GM/c^2 \approx R \tag{13}$$

Thus, the fact hat the gravitational radius of the Universe is of the same order as its size should no longer be considered as a coincidence.

Comparing (10) and (13) we get the value of w as

$$W = -1/3 \quad \text{so that} \quad p = -1/3 \, \rho \tag{14}$$

We note that this is the same as the equation of state for photons, with a minus sign. We interpret it as the negative pressure of the gravity quanta.

The two solutions found correspond to two different points of view (observers). The solution with w = 0, and an expanding Universe, is the local point of view with zero pressure, zero non-localized gravitational energy or gravity quanta pressure, and a cosmological constant different from zero. The solution with w = -1/3 is the cosmic point of view, the "cosmic observer" that sees gravity quanta pressure of the same type as the

250 photon gas pressure, withy negative sign that corresponds to gravity, and
251 zero cosmological constant.

252

253 **5. - WEINBERG'S RELATION**

254 In 1972 Weinberg discussed a relation obtained by using G
255 (gravitation), c (relativity), ℏ (quantum mechanics) and the Hubble
256 parameter H (cosmology) to arrive at the mass m of a fundamental particle:

$$m^3 \approx \frac{\hbar^2 H}{Gc} \qquad (15)$$

258 This is an intriguing relation that we will validate: classically all
259 parameters in it but one (H) is time varying¡ It is also an expression that
260 connects local parameters with a cosmological one: the Hubble H. Our
261 analysis will prove that, whatever the possible time variations of G and m,
262 ℏ being a universal constant, the speed of light c must be proportional to H
263 and therefore time varying from a cosmological point of view  This leaves
264 Weinberg's relation (15) with no time dependence, fully explained, and
265 therefore should no longer be considered as a coincidence¡

266 We will follow a Machean approach to validate this relation (and
267 using the laws of gravitation and the conservation of linear momentum) by
268 considering that the maximum momentum content of a mass m is mc, a
269 constant due to the presence of the rest of the Universe. It has acquired this
270 momentum during a time t. Then the total force that could  exert is mc/t .

The same result of this heuristic argument can also be obtained by considering the rate of change with time of the total relativistic energy of the mass M, $Mc^2$, divided by c and considering the universal constant Mc:

$$\frac{1}{c}\frac{dMc^2}{dt} = M\frac{dc}{dt} \qquad (16)$$

We later will prove that the speed of light c is proportional to the Hubble parameter H, c = HL, where L is a constant. For an expanding Universe the time derivative of H is of the order of $-H^2$ so that in absolute value we have for (16)

$$M\frac{dc}{dt} = MH^2L = MHc \qquad (17)$$

This total intrinsic force power (17) can mechanically be thought of as fluxing over a spherical surface of radius r centred on the centre of gravity of M. It will act upon the small area represented by m, m << M, that has a size of the order of its Compton wavelength. Then m will feel a force exerted by the presence of M and given by

$$\frac{GMm}{r^2} = MHc\frac{(\hbar/mc)^2}{4\pi r^2} \qquad (18)$$

and rearranging we arrive at

$$4\pi m^3 = \frac{\hbar^2 H}{Gc} \qquad (19)$$

which is the Weinberg's relation (15) within a factor of less than 3. From our astrophysical constants, $c^3/G$ and mc, we arrive at

$$c = H \times \text{constant} = H L \qquad (20)$$

This conclusion has already been presented elsewhere, Alfonso-Faus, 2007. Substituting (20) into (15) we get the new Weinberg's relation

$$m^3 \approx \frac{\hbar^2}{GL} \qquad (21)$$

Here we see that we have obtained a classical relation with G, m, ℏ and L all constants, and no time dependence is explicit. But leaving G and m possibly to vary with time we have, for ℏ and L as true constants, the new constant

$$Gm^3 = \text{constant} \approx \hbar^2/L \qquad (22)$$

This is what is observed in the lunar range experiments with laser. Since no time variation is assumed in m by the researchers, no time variation is observed in G.

## 6. - INTEGRATION OF THE EINSTEIN COSMOLOGICAL EQUATIONS

The two cosmological equations are

$$\left(\frac{\dot{a}}{a}\right)^2 + \frac{2\ddot{a}}{a} + 8\pi G \frac{p}{c^2} + \frac{kc^2}{a^2} = \Lambda c^2 \qquad (23)$$

307
$$\left(\frac{\dot{a}}{a}\right)^2 - \frac{8\pi}{3}G\rho + \frac{kc^2}{a^2} = \frac{\Lambda c^2}{3} \qquad (24)$$

308 where we have use the notation *a*, the cosmological scale factor, instead of
309 the previous R. It is well known that the equation resulting from the
310 integration of the Bianchi identity, the conservation equation (7) or (8), can
311 be considered with one of the (23) or (24) as the complete set of Einstein's
312 cosmological equations. We repeat here its integration (11) with the new
313 nomenclature,

314
$$a(t)^{3w} = const \qquad (25)$$

315 If the Universe is expanding $a$(t) is some function of time and w must be
316 zero. Some time ago the cosmological equation (23) was normally
317 considered to have the pressure term as due to the thermal motion of
318 galaxies, the Universe considered as a gas of galaxies. And the usual value
319 for this pressure was zero, i.e. w = 0. Our result demands that, should there
320 be any expansion at all, w must be exactly zero. We will analyze this case
321 with the solution $a$(t) = At$^2$ that has an accelerated expansion built in. Then
322 equation (23) gives

323
$$2\left(\frac{2}{t}\right)^2 + \frac{kc^2}{a^2} = \Lambda c^2 \qquad (26)$$

324 and equation (24) is now

$$\left(\frac{2}{t}\right)^2 - \frac{8\pi}{3}G\rho + \frac{kc^2}{a^2} = \frac{\Lambda c^2}{3} \qquad (27)$$

These two equations have the following expressions, using the normal cosmological nomenclature of dimensionless omegas:

$$2 + \Omega_k = 3\Omega_\Lambda \qquad (28)$$

$$1 - \Omega_m + \Omega_k = \Omega_\Lambda \qquad (29)$$

Today observations favour a flat Universe $\Omega_k = 0$, so that (28) gives $\Omega_\Lambda = 2/3$ and from (29) $\Omega_m = 1/3$, which are the numerical values observed today too. We conclude that the solution $a(t) = At^2$ is well within the results measured today for $\Omega_k$, $\Omega_m$ and $\Omega_\Lambda$. It also gives an accelerated expansion with the acceleration parameter q = - ½ . The value of w = 0 is interpreted as being due to the zero value of the gravitational energy when locally observed. Also taking the galaxies as atoms of the Universe, in a kinetic theory point of view, their "pressure" is zero.

Let us see the case with w different from zero. This must correspond to a "cosmic observer", a non-local observer with a cosmological road. The gravitational energy does not have to be zero now. The cosmic observer sees the gravity quanta. In fact we have already obtained the value of w in (14) as w = -1/3. With such a non zero value there is no escape: equation (25) forces to consider that the "cosmic observer" sees no expansion at all. With $a(t)$ = constant the two cosmological equations (23) and (24) are now

$$-\Omega_m + \Omega_k = 3\Omega_\Lambda$$

$$-\Omega_m + \Omega_k = \Omega_\Lambda \quad (30)$$

This set of equations inevitable gives for the cosmic observer:

$$\Omega_\Lambda = 0$$

$$\Omega_m = \Omega_k \quad (31)$$

The result is that cosmically there is no expansion, no dark energy: they are local effects. Given the second equation in (31), it is consistent to interpret that cosmically there is curvature (K = 1) and that the two parameters for curvature and mass are the same, may be very small, of the order of 0.03. No dark matter would be needed in this model unless we have a higher value for the curvature parameter. The local view is different: no curvature (K = 0), no pressure (w = 0) and the values $\Omega_m = 1/3$ and $\Omega_\Lambda = 2/3$, as observed. In this case there is expansion at a rate $a(t) \propto t^2$.

## 7. - THE INITIAL CONDITIONS OF THE UNIVERSE

During the first instants of time the speed of light must have been almost constant. This means that the Hubble parameter then was constant so that we have at that time:

$$H = a'/a = \text{constant} \quad (32)$$

Integrating the above relation we get the solution:

$$a = a_0 e^{Ht} \qquad (33)$$

This exponential expansion is equivalent to the very well known inflation phase at the initial stages of the Universe. During the inflation phase the first Planck's fluctuation of size $10^{-33}$ cm expanded very rapidly to the size $10^{28}$ cm, as of today's Universe, so that we have

$$e^{Ht} = 10^{61} \qquad (34)$$

If $t_1$ is the first tic of the Universe, and $t_i$ is the duration of the inflationary phase, then

$$t_i/t_1 = Ht = 61 \ln 10 \approx 140 \qquad (35)$$

Hence, only 140 tics were necessary to inflate the Planck's fluctuation to the present size of the Universe, and thereafter it remained at constant size. This is the cosmic view.

## 8. - EXPLANATION OF THE PIONEER 10/11 ANOMALOUS ACCELERATION

The fact that the speed of light varies with time as in (20) has enormous cosmological implications. We can get the photon acceleration as

$$dc/dt = L \, dH/dt \qquad (36)$$

Since H is a'/a we get for the photon acceleration $a_f$

$$a_f = dc/dt = - L \, H^2 = - Hc \qquad (37)$$

Taking a range of values for H from 60 to 75 Km/sec/Mpc, the theoretical acceleration (37) for the photons is

$$a_f = -(7.13 \pm 0.3) \text{ cm/seg}^2 \quad (38)$$

The observed anomalous acceleration in the Pioneer 10/11 case, Anderson et al. 1998, is

$$a_p = -(8.74 \pm 1.33) \text{ cm/seg}^2 \quad (39)$$

which is within the theoretical margin given in (38). From this point of view the observed acceleration is not satellite acceleration, it is not of gravitational origin on the satellite. It is in the photons as "cosmic observers" because their speed varies with cosmological time.

## 9. – THE COSMOLOGICAL LINE ELEMENT WITH THE NEW RESULTS. THE FUNDAMENTAL PROPERTY OF CLOCKS.

We use a Robertson-Walker metric as follows, with $ds$ the line element

$$ds^2 = a^2(t)\left(\frac{dr^2}{1-kr^2} + r^2 d\theta^2 + r^2 \sin^2\theta \; d\phi^2\right) - c^2 dt^2 \quad (40)$$

where t is the cosmic standard time, or a function of it, as defined by Weinberg, 1972. We are now ready to see that all clocks have a tic-tac duration proportional to t. The atomic clocks have a tic-tac proportional to $\hbar/(mc^2)$ = constant/c. Since we have proved that the speed of light is proportional to H, the inverse $1/c$ is proportional to t. The acceleration of

408  gravity is equal to GM/r² so that the period of any pendulum is inversely
409  proportional to $(GM)^{1/2}$ i.e. proportional to 1/c, the same thing. So is the
410  period of planets around the sun. Planck's time has the same property. All
411  clocks have the fundamental cosmological property that their tic-tac
412  duration is proportional to t. Then the product of any speed, which is
413  proportional to 1/t, by the tic-tac duration, is a characteristic constant
414  length. For cosmology, the whole Universe, this length is ct = constant = L
415  $\approx 10^{28}$ cm.

416  The time interval represented by one tic-tac *dt* in (40) is then
417  proportional to t so that cdt is equal to Ldt´, where dt´is the numbering of
418  tic-tac from a certain origin. Since cosmologically is *a(t)* = constant = L,
419  equation (40) transforms to

$$ds^2 = L^2\left(\frac{dr^2}{1-kr^2} + r^2 d\theta^2 + r^2 \sin^2\theta \; d\phi^2\right) - L^2 dt^2 \quad (41)$$

421  or in a dimensionless form

$$\left(d\frac{s}{L}\right)^2 = \left(\frac{dr^2}{1-kr^2} + r^2 d\theta^2 + r^2 \sin^2\theta \; d\phi^2\right) - dt^2 \quad (42)$$

423  where *s/L* is the percentage of the length of the line element *s* with respect
424  to the size of the Universe L. We see that the coordinate length *r* is
425  dimensionless, the same as dt´. They are pure numbers defining the
426  coordinate points in space and time.

427 We have seen that locally *k = 0* is a very good approach giving the
428 standard observations of $\Omega_m$ = 1/3 and $\Omega_\Lambda$ = 2/3. Hence the local line
429 element is given by (taking L = 1)

$$ds^2 = dx^2 + dy^2 + dz^2 - dt'^2 \tag{43}$$

431 which is the Minkowski space with the speed of light c = 1. On the other
432 hand at a cosmological level, as seen by a cosmic observer like a free
433 photon, we have *k = 1* and the cosmic line element is now, L = 1,

$$ds^2 = \left(\frac{dr^2}{1-r^2} + r^2 d\theta^2 + r^2 \sin^2\theta \, d\phi^2\right) - dt^2 \tag{44}$$

435 This spherical metric has no infinites. When r is approaching 1 the radial
436 element has a finite limit, where $1 - r^2$ tends to zero as well as $dr^2$. This is
437 the metric as seen by a photon (*ds = 0*) with speed c = 1 at a distance from
438 the origin *r* = 1 = L with the age of the Universe t taken as just one tic (*dt'*
439 = 1), c = 1/t. This is the expression of the speed of light, c = 1/t, when we
440 take L = $10^{28}$ cm = 1, i.e. ct = 1.

## 10. – CONCLUSIONS

443 We have found two long sought explanations for the two
444 "coincidences": the gravitational radius of the Universe equal to its size,
445 both constants in our treatment, and the Weinberg's relation that defines a
446 fundamental mass in terms of a cosmological term, the Hubble parameter

H. This parameter disappears when the proportionality of the speed of light c with H is introduced in the relation. The anthropic principle is not necessary.

With the application of a field to particle technique we have proved that all gravitational radii are constant. Using this result we have integrated the Bianchi identity, and the Einstein cosmological equations, which show two solutions for the Universe: one local, flat accelerated expanding Universe with non-zero cosmological constant, predicting the numerical values already observed for the omega parameters ($\Omega_k = 0$, $\Omega_\Lambda = 2/3$, $\Omega_m = 2/3$, w = 0, $a(t) \propto t^2$). This is the standard cosmological model. Therefore there is an obvious complete agreement with the present status of cosmology. There is an initial inflation period that converts one Planck's fluctuation size into an inflated Universe of the present size in about 140 tic-tac.

The other solution is interpreted here as the cosmological one, a new solution found by integrating the Bianchi identity, ( $a(t)$ = constant), and the cosmological equations ($\Omega_\Lambda = 0$, k = 1, $\Omega_k = \Omega_m$, w = -1/3). It is the point of view of a cosmic observer, as for example a free photon, which sees a curved spherical Universe with zero cosmological constant. There is no need for either dark matter or dark energy at this scale. We validate Weinberg's relation and find that the speed of light c is proportional to the Hubble parameter. This time-varying situation for c predicts something already observed, with no explanation as of to day. The free photons coming from the Pioneer 10/11 spacecrafts exhibit an acceleration that is well within the predicted value. Thus the anomalous acceleration of these spacecrafts is fully explained, and no anomalous acceleration in them exist coming from any gravitational origin, a conclusion that is abundant in the scientific literature. The key point is the slowing down cosmological law for c and any v, proportional to $1/t$. This also lowers the temperature. The growing gravity due to the increase of mass makes the initial plasma (after inflation) to condensate into a proto galaxy. This was the first reason to postulate the existence of "dark matter" that is not necessary here.

Not only photons are cosmic observers. There are also neutrinos, tachyons and gravity quanta. We expect more predictions from this cosmological solution deduced from the integration of the field equations.

## 11. – REFERENCES